\documentclass[twocolumn,prl,superscriptaddress,amsmath,amssymb,aps]{revtex4-1}

\makeindex

\usepackage{xcolor}  
\usepackage{appendix}
\usepackage{float}  
\usepackage{placeins}
\usepackage{graphicx}
\usepackage[percent]{overpic}
\usepackage{bm}

\begin{document}

\title{The Structure of Spreading on Temporal Networks}

\author{Omar Henderson}
\affiliation{Department of Network and Data Science, Central European University, 1100 Vienna, Austria}

\author{Mikko Kivelä}
\affiliation{Department of Computer Science, Aalto Univerisity, 00076 Espoo, Finland}

\author{Márton Karsai}
\affiliation{Department of Network and Data Science, Central European University, 1100 Vienna, Austria}
\affiliation{HUN-REN Rényi Institute of Mathematics, 1053 Budapest, Hungary}


\begin{abstract}
The physics of spreading in static networks is well understood through mappings to percolation. We show that spreading dynamics on temporal networks can analogously be mapped to reachability in temporal event graphs. This provides a theoretical and computational framework for a class of processes, such as variants of the susceptible–infected–susceptible model. Without explicit simulations, through the component analysis of event graphs, we obtain epidemic prevalence and derive epidemic thresholds for temporal networks with arbitrary degree and inter-event time distributions, with significant computational advantages as compared to explicit simulations.
\end{abstract}

\maketitle

Spreading is among the most fundamental processes occurring on networks: a pathogen, a piece of information, or an adopted behaviour either dies out or invades a finite fraction of the system. Spreading processes are typically understood as critical phenomena, in which the threshold between the vanishing and spreading phases, together with the invaded fraction, are the central objects of theoretical and practical interest. 
As early as a century ago~\cite{kermack1927contribution}, the fundamentals of spreading processes were understood through idealised models of well-mixed populations, often with memoryless transitions, allowing the dynamics to be reduced to systems of differential equations. Much later, network theory traded the well-mixed assumption for an explicit static contact topology, revolutionising our understanding of how structure affects spreading~\cite{pastor2015epidemic}. 
Crucially, this analytical power was retained: for specific processes, the final epidemic size and epidemic threshold can be mapped onto percolation problems~\cite{newman2002spread,kenah2007second,boguna2013nature}, allowing spreading outcomes to be computed from network connectivity rather than from repeated simulation.

Real-world systems, however, are seldom static and are more accurately characterised by time-varying interactions represented as temporal networks~\cite{holme2012temporal,LambiotteMasuda2016GuideTemporalNetworks}. The time-varying interaction patterns can fundamentally alter any ongoing dynamical processes and lead to radically different outcomes, as demonstrated in several systems over the last decade~\cite{LambiotteMasuda2016GuideTemporalNetworks,Karsai2011slowsmallworld,holme2014birth,MasudaHolme2013F1000,lee2012exploiting,holme2012temporal,kivela2012multiscale,stehle2011simulation}. These studies show that bursty contact patterns and event durations, together with the causal ordering of interactions and their correlations with the underlying static structure, can significantly impact the outcome of many spreading phenomena~\cite{Karsai2011slowsmallworld,HorvathKertesz2014,holme2014birth,MasudaR0,onaga2017concurrency}.

Nevertheless, while the importance of time-varying interactions for spreading processes has been recognised, most studies rely on observations obtained from stochastic numerical simulations, as few effective analytical tools are available for handling the complexity of temporal networks ~\cite{Karsai2011slowsmallworld,holme2014birth,rocha2011sexbursty,MasudaHolme2013F1000}. 
Such approaches are effective because they are agnostic to how temporal and static structures are represented. However, due to this generality, they lack direct insights into which structures govern spreading, while also being computationally expensive. In contrast, analytical solutions for static networks can exploit structural information to compute the final epidemic size directly from the connectivity of the underlying network~\cite{newman2002spread}.
The few analytical studies that have been proposed for temporal networks either rely on spectral methods based on discrete-time supra-adjacency matrices~\cite{valdano2015threshold}, extend these to continuous time under specific conditions~\cite{valdano2018epidemic}, or build on schematic random network models like the activity driven framework~\cite{Perra2012ActivityDriven,karsai_time_2014,Mancastroppa2019Burstiness}, which capture only a limited number of temporal network characteristics.

In this paper, we establish a link between spreading processes and temporal network connectivity that offers conceptual and computational advantages analogous to those between spreading and percolation in static networks~\cite{newman2002spread, pastor2015epidemic}. We use event graphs (EGs) to represent temporal networks as static directed acyclic graphs (DAGs)~\cite{mellor_teg,saramaki_wteg}, encoding the superposition of all time-respecting paths, which serve as substrates for any dynamical process. Recent work has shown that the connectivity transition of EGs belongs to the directed percolation universality class ~\cite{badie2022directed,badie2022directedhet}, allowing spreading on temporal networks to be studied using established directed percolation theory~\cite{hinrichsen2000non}.

We show that a single EG realisation can simultaneously capture the spreading process structure starting from any initial conditions for globally consistent spreading processes. 
As canonical examples of such processes, we introduce variants of Susceptible-Infected-Susceptible (SIS) processes, which map exactly onto percolation on the event graph where the infection probability corresponds to a site percolation problem and the recovery time to bond percolation.

\begin{figure*}[!t]
\centering
\includegraphics[width=0.8\linewidth]{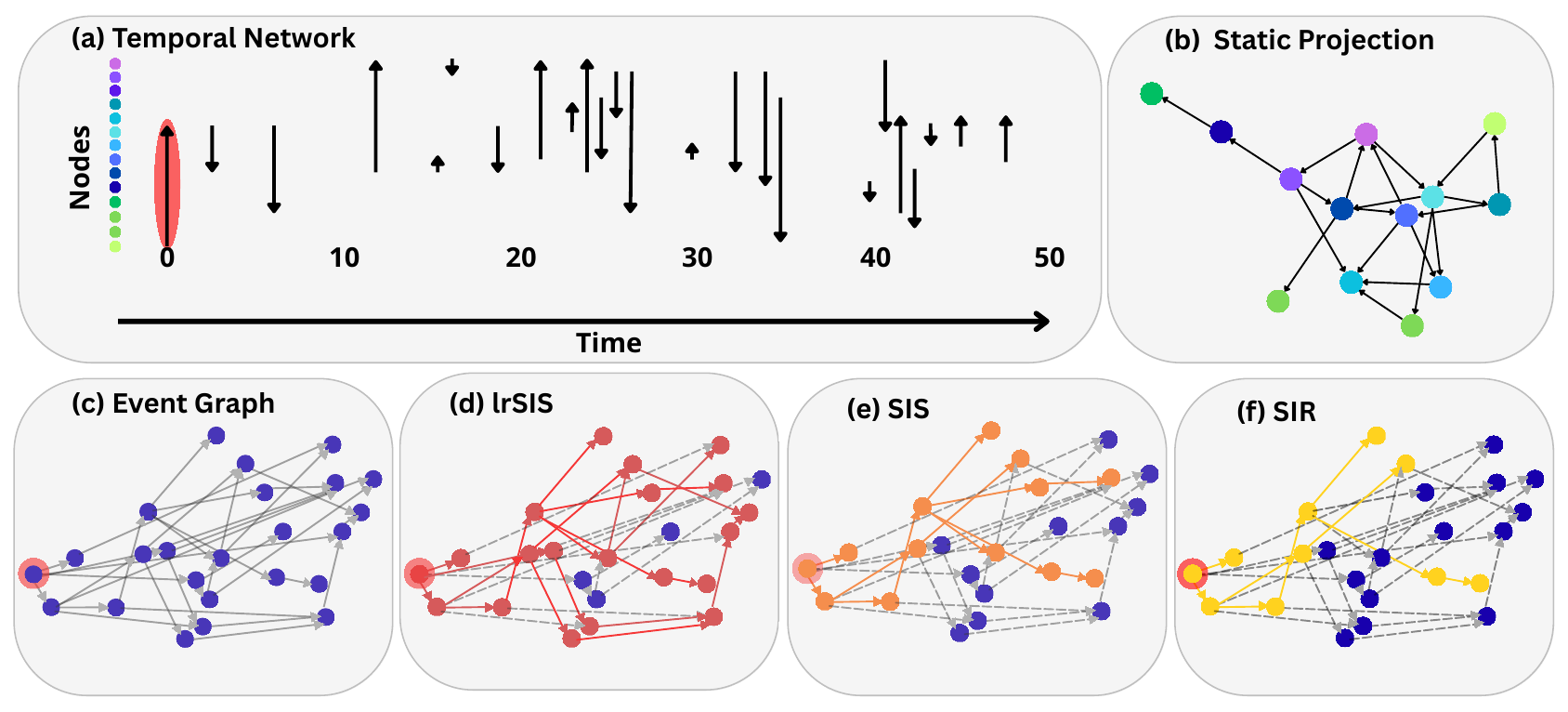}
\caption{
Schematic of (a) a temporal network, (b) its static projection, and its (c) event graph representation. Spreading paths starting from the first event (red circle) as seed are shown for the (d) lrSIS/frSIS (red), (e) SIS (orange), and (f) SIR (yellow) processes. Each process uses $\beta=1, \delta t=15$.}
\label{fig:Schematic}
\end{figure*}
Furthermore, we derive an analytical threshold condition for random temporal networks with arbitrary degree and inter-event time distributions, and explore how static and temporal interaction patterns jointly influence the structure of spreading in modelled and empirical networks. We demonstrate the computational advantages of our approach, similar to the static case \cite{newman2002spread}, where instead of expensive explicit simulations, a relatively cheap operation can provide the connected components of an appropriately pruned temporal event graph.

A directed temporal network $G=(\mathcal{V},\mathcal{E},\mathcal{T})$ consists of a set of vertices $\mathcal{V}$, and a set of time-stamped events $e=(u,v,t^{start},t^{end})\in \mathcal{E}$ describing directed events between a source $u$ and a target node $v$ starting at time $t^{start}$ and ending at $t^{end}$, both in $ \mathcal{T}$ ~\cite{badie2022directed}. For a schematic temporal network see Fig.\ref{fig:Schematic}(a), and for its aggregated static structure (b). The event graph representation relies on the notion of temporal adjacency, denoted $e_i\rightarrow e_j$, between two directed events $e_i,e_j\in \mathcal{E}$. Two events are temporally adjacent if the second follows the first chronologically ($t_i^{end}<t_j^{start}$) and $v_i = u_j$, i.e. the target of the earlier event matches the source node of the later event ~\cite{mellor_teg,saramaki_wteg}. For simplicity, in the following we assume events are directed and instantaneous (i.e. with $t_i^{start}=t_i^{end}$, marked as $t_i$) unless stated otherwise. Our results can be generalised for undirected events (see Appendix~\ref{app:undirected}) and with duration.

A \emph{temporal event graph} (EG) is  a directed static network $D=(\mathcal{E},E_D,\Delta t(e_i,e_j))$, whose nodes are the events of the original temporal network $G$, and directed links $(e_i,e_j,\Delta t(e_i,e_j))\in E_D$ encode temporal adjacency between events, with direction respecting their temporal order (as shown in Fig.\ref{fig:Schematic}(c)). Each link in $E_D$ is assigned a weight, equal to the waiting time between the two adjacent events $\Delta t(e_i,e_j)=t_j-t_i$. Temporally adjacent events, satisfying $\Delta t(e_i,e_j)<\delta t$ are called $\delta t$-adjacent, representing a path where a limited waiting-time process, e.g. a spreading process with finite recovery time $\delta t$, could pass through. Hence,  constructing the event graph of a temporal network, yields a directed static weighted acyclic graph, which encodes all  time-respecting paths ~\cite{badie2020efficient}. By removing links from the event graph with weights $\Delta t \geq \delta t$, we retain only the time-respecting paths that can sustain a spreading process with waiting-time limit $\delta t$. This leads to a bond percolation process with $\delta t$ as control parameter, mapping several random temporal networks into the directed percolation universality class
~\cite{badie2022directed,badie2022directedhet}.

The paths a spreading process can take on a temporal network are limited to time-respecting paths, which map into the static paths in the corresponding EG ~\cite{mellor_teg,kivela2018mapping}. Thus, the path of a spreading process corresponds to a subset of EG nodes, which are reachable from the source node. Although several paths may exist between two nodes, adjacency relations can be pruned to recover the spreading routes of any epidemic process, 
as illustrated for SIS and SIR processes (where R stands for Recovered) in Fig. \ref{fig:Schematic} (e) and (f), respectively. However, recovering these spreading structures while retaining analytical tractability and the corresponding computational advantages, is where the challenge---and opportunity---presents itself. While all spreading processes are embedded in the EG, some can be fully coded by a single EG structure, while for others, the EG structure depends on the initial conditions.

We call a spreading process \emph{globally consistent} if all potential realisations can be represented as reachability on a single event graph, independent of the initial conditions and process history. Processes for which the spreading structure depends on the initial conditions are \emph{globally inconsistent} and may require a different EG pruning for each realisation.

This distinction reflects the compatibility between the contact pattern and the dynamical process defined on it. In well-mixed populations, the compatible choice is Markovian: exponentially distributed recovery times render the dynamics memoryless and reduce it to differential equations~\cite{kermack1927contribution}. While this assumption is often carried over to network epidemic models~\cite{newman2002spread}, on static networks the compatible choice is actually a fixed infectious period, for which the SIR process is exactly isomorphic to a percolation problem~\cite{kenah2007second}. On temporal networks, the compatible choice is global consistency: recovery must respect EG adjacency, so that the process is isomorphic to reachability on a single event graph.

An SIS process is a natural candidate for compatibility with the EG structure, motivated by its well-established connection to directed percolation~\cite{hinrichsen2000non,Grassberger1983,CardySugar1980,Mata2021}. However, standard variants of the SIS process, with Markovian transitions between states or where an infected node recovers after a fixed period $\delta t$ following the initial infection event, are not globally consistent. Instead, if the infectious period of an infected node is reset after every incoming event from an infected neighbour, the process becomes globally consistent. We call this the last-contact reinforcing SIS (lrSIS) process, as the fixed recovery time is always computed from the last infecting event in a $\delta t$ adjacent interaction sequence of an infected node. This definition represents processes where a nodes infectious period expands due to repeated infections. The lrSIS maps exactly to reachability on the event graph for fixed recovery time $\delta t$ (see Fig.~\ref{fig:Schematic}(d) for an example and Appendix~\ref{app:rsis-eg-equivalence} for proof).
Note that, beyond analytical convenience, the detailed recovery mechanisms of SIS models are application-dependent modelling choices: although reinforcement can be unnatural for pathogens, it is natural for opinion and behaviour spreading, where repeated exposure sustains adoption.

Despite the different recovery mechanisms, the lrSIS and the more standard SIS processes with fixed recovery time but no reinforcement (what we refer to as the SIS process from here on throughout the paper) are closely related: for any temporal network and seed, the number of infected nodes in an lrSIS process is an upper bound for the same quantity in a corresponding SIS process (see Appendix \ref{app:rsis-upper-bound} for proof). Nevertheless, global consistency of the lrSIS process enables significant analytical and computational advantages. As it is isomorphic to reachability on a static DAG, it allows analytical tractability of spreading dynamics and the derivation of spreading outcomes from out-component analysis in the corresponding DAGs, with cheap computational solutions~\cite{badie2020efficient}, instead of costly numerical simulations or simplified analytical approximations.

We further introduce the first-contact reinforcing SIS (frSIS), a globally consistent variant that more closely resembles SIS with fixed recovery time. As in lrSIS, incoming infectious interactions reset the recovery clock, however, an infected node can infect each neighbour at most once before its recovery clock is reset. Once reset, it may infect each neighbour again. This process maps to a constrained event graph $D_C=(\mathcal{E},E_C)$, where if $(e_i \rightarrow e_j)\in E_C$, then $\nexists (e_i \rightarrow e_k)\in E_C$ with $e_j$ and $e_k$ occurring on the same static link and $t_i<t_k<t_j$.

\begin{figure}[ht!]
    \centering

\includegraphics[width=.9\linewidth]{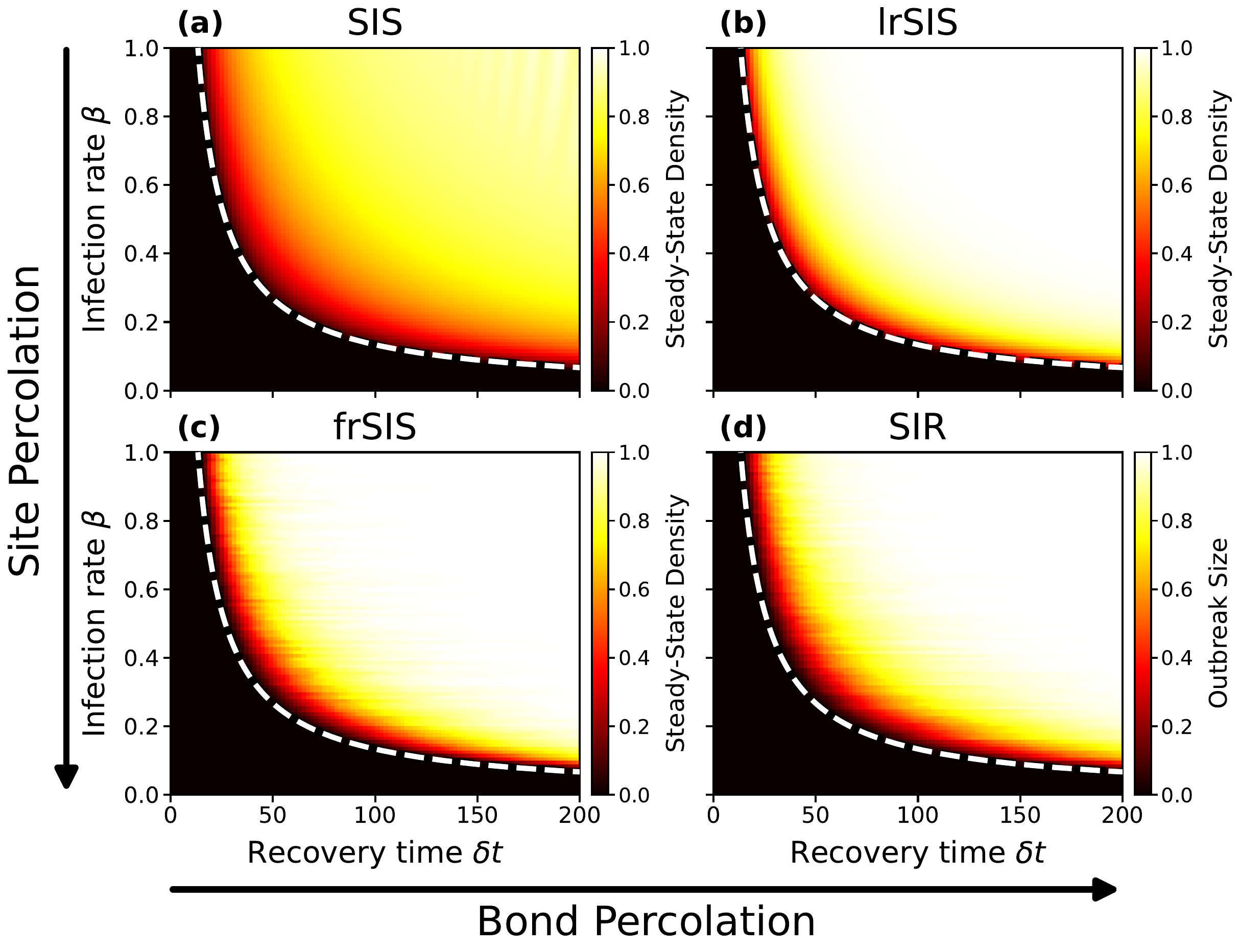}
    \caption{Phase diagrams in $(\delta t,\beta)$ for (a) SIS, (b) lrSIS, (c) frSIS processes and (d) SIR on k-regular temporal networks with $k=8$ and Poisson link activation  rate $\lambda=0.01$. (a) and (d) are obtained from simulations, while (b) and (c) are derived from EG component analysis. Dashed line shows critical threshold derived from Eq.~\ref{eq:poissondtc}.}
    \label{fig:bond_site}
\end{figure}

Both bond and site percolation on the event graph map to meaningful phenomena in the underlying spreading dynamics.
The recovery time $\delta t$, which thresholds the event graph links, is the bond percolation parameter, while the infection probability $\beta$ defines site percolation by removing events that fail to propagate the infection. Figure~\ref{fig:bond_site} shows the corresponding phase diagrams for SIS (a), lrSIS (b), frSIS (c), and SIR (d) on $k$-regular temporal networks with Poisson link activation. All systems exhibit similar epidemic transitions between a susceptible and infected phase, characterized by a transition contour that appears very similar between each spreading process.

Since the lrSIS and frSIS  can be reframed as reachability problems on the event graph, their spreading behaviour can be analysed through its structural properties. In particular, if the EG is locally tree-like, spreading can be described as a branching process. By further assuming random temporal link dynamics, generating functions  can be used to derive the degree distribution of the emerging EG. We therefore approximate the expected excess out-degree of the EG, which is analogous to the basic reproduction number $R_0$ in spreading processes~\cite{newman2002spread}. The generating function depends on both the structure of the underlying network and the activation probability of each link. Therefore, for random directed graphs, the composite generating function of the event graph takes the form:
\begin{equation}
G(x,y) = \sum_{k_{in},k_{out}}p_{k_{in}k_{out}}x^{k_{in}}H(y)^{k_{out}},
\label{eq:PGF}
\end{equation}
where $k_{in}$ and $k_{out}$ are the in- and out-degrees of the underlying static network (respectively), and $H(y) = (1-p)+py$ is the generating function for the activation of a single edge with probability $p$. Since event times are independent between links, infections arrive at a random point in a link's timeline ~\cite{Karsai2011slowsmallworld,kivela2015estimating}. Hence, $p$ is given by the residual waiting time CDF $F_R(\delta t)$ ~\cite{kivela2015estimating}. 
Thus, for temporal networks with independent events over an underlying directed configuration model static network, $R_0$ can be derived from Eq.\ref{eq:PGF} as:
\begin{equation}
    R_0=F_R(\delta{t})R_0^{stat},
    \end{equation}
where $R_0^{stat}= \frac{\langle k_{in}k_{out}\rangle}{\langle k_{in}\rangle}$ is the basic reproduction number in a static directed network and $\langle k \rangle=\langle k_{in} \rangle =\langle k_{out} \rangle $.
Hence, the epidemic threshold ($R_0=1$) occurs when
\begin{equation}
F_R(\delta{t_c})={R_0^{stat}}^{-1}.
\label{eq:Condition Solution}
\end{equation}
For the full derivation, see Appendix~\ref{app:Derivations}. Also note, that if the activation probability $p$ depends on $k_{in}$ or $k_{out}$, then the threshold condition takes a different functional form.
\begin{figure}[!t]
    \centering
\includegraphics[width=0.9\linewidth]{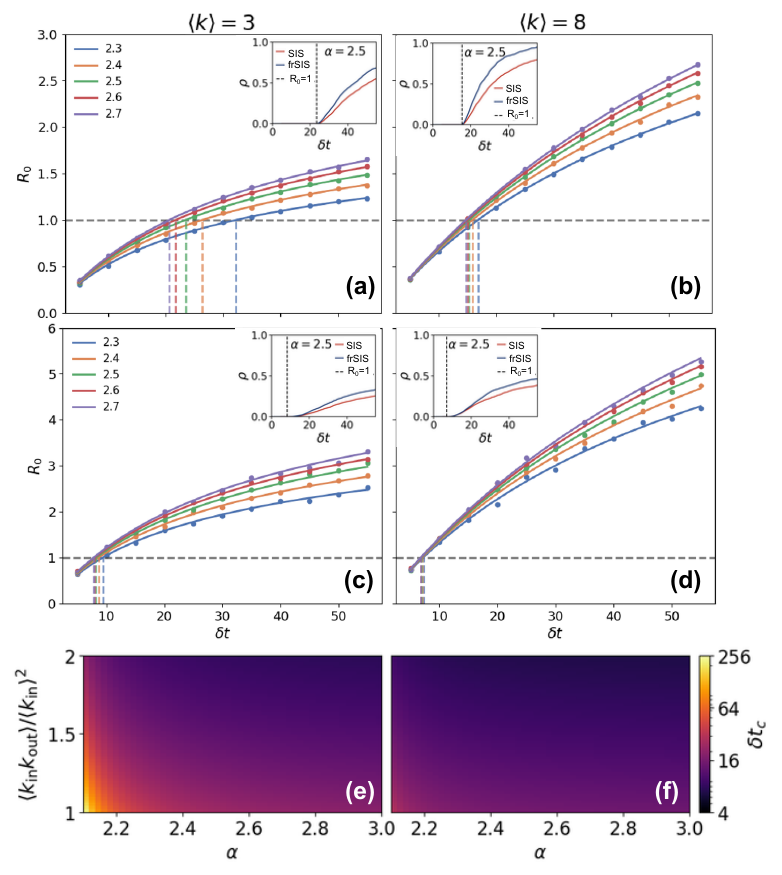}
    \caption{Basic reproduction number $R_0$ as function of recovery time $\delta t$ and inter-event time exponent $\alpha$ for temporal networks with bursty link dynamics. Solid lines show analytical predictions, symbols numerical results and dashed lines show the thresholds $R_0=1$. Left panels correspond to sparse ($\langle k\rangle=3$) and right to dense ($\langle k\rangle=8$) networks. Panels (a, b) show $k$-regular networks, while (c, d) degree-heterogeneous networks with $\langle k_{in}k_{out}\rangle \approx2\langle k\rangle^2$. Mean inter-event times are $\langle\tau\rangle=37.5$ (left) and $100$ (right), preserving the overall and node-specific activity rates. Insets compare SIS and frSIS steady-state densities, showing coincident epidemic onsets, with mean-field predictions more accurate for homogeneous networks. Panels (e,f) show analytical $\delta t_c$ for varying degree and inter-event time distributions.}
    \label{fig:R0}
\end{figure}    

As the simplest interaction dynamics, let's assume that times of consecutive events on a link are generated by a Poisson process with rate $\lambda$, and a residual waiting time distribution $F_R(\delta{t})=F(\delta{t})=1-e^{-\lambda\delta{t}}$.
Solving Eq.~\ref{eq:Condition Solution} in this case, the critical recovery time appears as:
\begin{equation}
\delta{t}_c=-\frac{1}{\lambda}\log\left(1-{R_0^{stat}}^{-1}\right),
\label{eq:poissondtc}
\end{equation}
which is precisely the epidemic threshold condition for a fixed recovery time SIS model on static networks \cite{Kiss2015NonMarkovianPairwise}. This is demonstrated in Fig.~\ref{fig:bond_site}, where the critical line, derived from Eq.~\ref{eq:poissondtc}, coincides well with the phase-transitions separating a susceptible and epidemic phase as the function of $\delta t$ and $\beta$. While the critical lines correspond, the processes diverge in the supercritical regime, where the lrSIS process forms an upper bound approximation of the more conventional SIS dynamics with fixed recovery time. This is because in directed networks, reinforcement requires a path of at least two events and hence only has a significant effect once the system is in the supercritical regime, when such paths become commonplace.

In more realistic bursty temporal networks, the static network approximation is no longer sufficient, as temporal heterogeneity affects the timing of events. To model bursty link dynamics we consider a Lomax distribution (also known as a Pareto Type II distribution)~\cite{johnson1994continuous} due to its flexibility to scale the characteristic time scale without imposing a hard cut-off, while keeping the average inter-event time $\mu$ fixed. The residual CDF of the Lomax distribution is
\[
F_R(\delta t)
=
1-\left(1+\frac{\delta t}{t_0}\right)^{-(\alpha-2)}, \delta t\geq0,
\], 
where $\alpha$ controls the tail exponent of the distribution and $t_0 = \mu(\alpha-2)$ is the characteristic timescale (for derivation see Appendix~\ref{app:Derivations}.

In this case, the critical point (at $R_0= 1$) is:
\begin{equation}
 \delta t_c
=
t_0\left[
\left({1-{R_0^{stat}}^{-1}}\right)^{\frac{1}{2-\alpha}}-1\right].   
\end{equation}
Since $F_R(\delta t) \to 0$ as $\alpha \to 2$, for any network with fixed static structure, the critical recovery time must diverge, $\delta t_c \to \infty$. In other words, as $\alpha \to 2$, the time-respecting paths, necessary for an epidemic to occur, exist only for increasingly long recovery times. This is demonstrated in Fig.~\ref{fig:R0}(a)-(d), which compares analytical and numerical estimates of $R_0$ for $k$-regular configuration model networks in panels (a) and (b) and for degree heterogeneous networks in panels (c) and (d). In each case, link activations follow a bursty renewal process with varying inter-event time exponent $\alpha$ (see colours). Strikingly, the analytical predictions (solid curves) agree closely with the simulation results (symbols) across all networks and parameter values. As expected, decreasing $\alpha$ -- hence increasing burstiness -- increases the critical waiting time $\delta t_c$. However this dependence is weak in dense and degree heterogeneous networks. This is also evident from the heat-maps in Fig.~\ref{fig:R0}(e) and (f), which shows the joint dependence of $\delta t_c$ on temporal heterogeneity quantified by $\alpha$ and structural heterogeneity quantified by $\langle k_{in}k_{out}\rangle/\langle k_{in}\rangle^2$. Effectively, dependency appears only in the degree homogeneous bursty limit in sparse networks (corresponding to the bottom left corner of Fig.~\ref{fig:R0}(e)). 

Note that in a fully susceptible population on a tree-like network, $R_0$ is identical for the frSIS and fixed recovery time SIS processes, becoming only an approximation in denser networks. Nevertheless, as shown in the insets of Fig.~\ref{fig:R0}, the quasi-stationary densities of SIS and frSIS exhibit a very similar epidemic onset and critical point, diverging only in the supercritical regime. This supports the event graph as an accurate approximation for locating the SIS epidemic threshold.

\begin{figure}[t]
    \centering
    \includegraphics[width=0.9\linewidth]{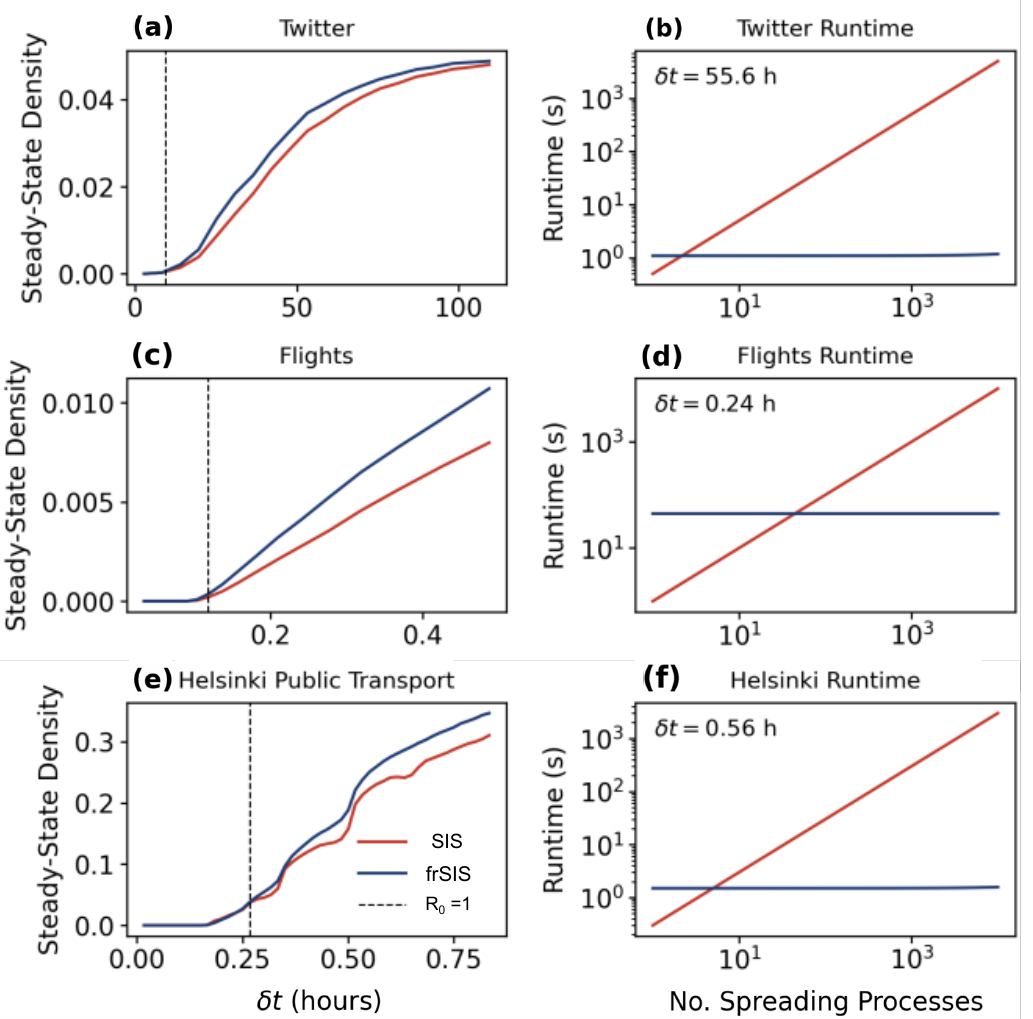}

        \caption{Steady-state infection size as function of $\delta t$ for fixed recovery time SIS and frSIS models on three empirical temporal networks: (a) Twitter ~\cite{yang2011patterns}, (c) global flights ~\cite{Olive2023OpenSky} and (e) Helsinki public transport~\cite{Kujala2018}. In all cases, frSIS forms a tight upper bound for the SIS. The numerical EG condition $R_0=1$ overlaid (dashed vertical line) show good agreement with the onset of epidemic in the Twitter and flight networks, but  overestimates the critical point of the Helsinki public transport network. Panels (b), (d) and (f) show the computational cost per simulation of explicit simulations and EG-based calculations for each network.}
    \label{fig:placeholder}
\end{figure}

To evaluate the validity of our framework in empirical systems, we focus on three empirical temporal networks: a network of replies on Twitter~\cite{yang2011patterns}, the global flight network for July 2019~\cite{Olive2023OpenSky}, and a typical day of the Helsinki public transport network~\cite{Kujala2018}. As these networks are not random, the assumptions underpinning the generating function approach do not generally hold, we therefore limit our exploration to numerical measurements. Epidemic outcomes in empirical temporal networks are highly sensitive to the choice of start time. To account for this, we sample observation windows from randomly selected starting times with duration T/64 for the Twitter network and T/2 for the other two networks. In each case we initialise the spreading process with a single infected seed at the beginning of the window and measure the average quasi-stationary state. As shown in Fig.~\ref{fig:placeholder}, in each of the empirical temporal networks, the agreement between the two processes around the threshold is remarkable. In addition, the numerical threshold condition $R_0=1$, measured directly from the EG, accurately predicts the onset of infection in the Twitter and Flight networks, while slightly overestimating the onset in the Helsinki public transport network. This discrepancy may arise from regions within the Helsinki public transport network, which are locally supercritical, even when the network is globally fragmented. To demonstrate the computational advantage of the EG method as compared to numerical simulations of the same spreading processes, we show the corresponding runtimes in Fig.~\ref{fig:placeholder}(b), (d) and (f), for a chosen $\delta t $ in the supercritical regime. Although event graph construction incurs an initial cost, once built, spreading outcomes for all initial conditions can be calculated in a single sweep \cite{badie2020efficient}. Hence as the number of realisations increases, globally consistent processes represented as EGs become much more computationally efficient. 

Spreading on temporal networks maps onto reachability in temporal event graphs in the same way that spreading on static networks maps onto percolation, and global consistency identifies the processes for which this mapping is exact. This correspondence is in one respect stronger than its static counterpart: percolation clusters give the final epidemic size, whereas event-graph out-components are ordered in time and therefore encode the full spreading history, including infection times and transmission routes. The consequence extends beyond the specific models with independent link dynamics treated here. In the past, percolation made complicated static network structures analysable, connecting degree heterogeneity~\cite{newman2002spread}, correlations~\cite{boguna2003absence}, clustering~\cite{miller2009percolation}, and more recently richer structural features~\cite{k2025homophily,hiraoka2025strength} directly to epidemic outcomes without recourse to simulation; event-graph reachability can do the same for complex temporal structures, as our derivation of $R_0$ for arbitrary degree and inter-event time distributions illustrates, and as the agreement with empirical networks supports. Further, because reachability delimits any dynamical process on a temporal network, it constrains complex contagion, random walks, diffusion, and opinion dynamics; wherever a globally consistent variant of these processes can be formulated, they too become structural problems.

\textbf{Acknowledgments} M.K. acknowledges support from the MOMA WWTF project; the BEQUAL NKFI-ADVANCED grant (No. 153172); and the COLINE DUT European Partnership project (F-DUT-2023-0037). M.Ki. acknowledge support from the Research Council of Finland (grant: 349366).

\textbf{Data Availability Statement}: All data used in this study are publicly available from the sources cited in the manuscript.

\textbf{Authors contribution}: All authors designed the research and wrote the manuscript. O.H. carried out the analytical and numerical calculations.


\providecommand{\noopsort}[1]{}\providecommand{\singleletter}[1]{#1}

\newpage

\appendix
\setcounter{secnumdepth}{1}
\begin{center}
\normalsize{\textbf{End Matter}}
\end{center}
\section{Event graphs vs. spreading processes}
\label{app:proofs}
\subsection{Equivalence between lrSIS process and event graph reachability}
\label{app:rsis-eg-equivalence}

\textbf{Statement:} The infected nodes in the lrSIS model directly map to the nodes in the out-components of an event graph with fixed recovery time equal to limited waiting time, and same starting seed. \\

\textbf{Proof:}

\emph{Step 1:} If an edge $(e_i,e_j)$ exists in the event graph then $t_j-t_i < {\Delta{t}}$ and the events must satisfy the relevant adjacency condition: for directed events, the target of $e_i$ is the source of $e_j$, while for undirected events, $e_i$ and $e_j$ must share at least one node. Since $e_i$ is reached, the node carrying the infection from $e_i$ to $e_j$ is infected at $t_i$. Therefore, since $t_j-t_i < {\Delta{t}}$, this node has not recovered before $t_j$ and $e_j$ is an infection event in the lrSIS process.

\emph{Step 2:} If an infection is carried from an event $e_i$ to event $e_j$ in the lrSIS model, then $t_j-t_i<\Delta{t}$ and the two events must be causally adjacent:$v_i=u_j$ for directed events or they must share at least one node for undirected events. This is precisely the condition for an edge to be present between two events in the event graph.

Hence, the presence of an edge in the event graph implies an infection path in the lrSIS and vice versa. Therefore starting from the same seed event, the set of events reached in the event graph is exactly the set of events activated in the lrSIS process.

\subsection{lrSIS process as an upper bound on fixed recovery time SIS process}
\label{app:rsis-upper-bound}
\textbf{Statement:} The size of an epidemic in an lrSIS process is an upper bound for a fixed recovery time SIS process starting from the same point, with the same recovery time.

\textbf{Proof:} If we assume that the SIS process reaches/infects more nodes than the lrSIS, then there must be at least one pair of events $(e_i,e_j)$ for which infection flows through in the SIS but not the lrSIS. If such an event pair exists in the SIS then this means that $t_j<t_i+ \delta{t}$ and there must be at least one shared node between $e_i$ and $e_j$. If these two conditions are met then $(e_i,e_j)$ must be an edge in the event graph and hence an infection path in the lrSIS model. Therefore, all pairs of events in the infection path in the SIS model must also exist in the lrSIS model.

\section{Derivation of $R_0$ basic reproduction number}
\label{app:Derivations}
Here we derive the expected number of secondary event-graph branches for the directed configuration model. Let $p_{k_{in}k_{out}}$ be the joint in- and out- degree distribution of the underlying static network, then the degree distribution can be expressed as 
\begin{equation}
G(x,y) = \sum_{k_{in}k_{out}}p_{k_{in}k_{out}}x^{k_{in}}H(y)^{k_{out}},
\label{eq:PGF_2}
\end{equation}
where $H(y)$ is the probability that a link has an event. For the link activation model, this can be written as 
$H(y)=1-F_R(\delta t)+F_R(\delta t)y$. Since the event times are independent across links, $F_R(\delta t)$ is the cumulative distribution function (CDF) of the residual inter event time distribution,
\begin{equation}
F_R(\delta{t})=\frac{1}{\mu}\int_0^{\delta t}{[1-F(u)]du}, 
\label{eq:Fr}
\end{equation}
where $F(u)$ is the CDF of the original inter-event time distribution and $\mu$ is its mean.

As, infections arrive via an incoming link, this biases the degree distribution, therefore we must take the excess out-degree distribution. The excess generating function can be written as
\begin{equation}
   G_1(y)
=
\frac{
\partial G(x,y)/\partial x \big|_{x=1}
}{
\partial G(x,y)/\partial x \big|_{x=1,y=1}
}. 
\end{equation}
Substituting Eq.~\eqref{eq:PGF_2} gives
\begin{equation}
    G_1(y)=\sum_{k_{in},k_{out}} \frac{k_{in}p_{k_{in},k_{out}}}{\langle k_{in} \rangle}H(y)^{k_{out}}.   
\end{equation}

The expected excess out-degree and hence the basic reproduction number is therefore obtained by
\begin{equation}
    R_0=G_1'(1)=F_R(\delta{t})\frac{\langle k_{in}k_{out}\rangle }{\langle k_{in}\rangle }
\end{equation}

In the case of bursty link dynamics following a Lomax distribution, the probability density function is 
\[
f(\delta t)
=
\frac{\alpha-1}{t_0}
\left(
1+\frac{\delta t}{t_0}
\right)^{-\alpha},
\qquad \delta t \ge 0,
\]
where $\alpha$ controls the tail exponent of the distribution and $t_0 = \mu(\alpha-2)$ is the characteristic timescale

The residual CDF of the distribution is therefore:
\[
F_R(\delta t)
=
1-\left(1+\frac{\delta t}{t_0}\right)^{-(\alpha-2)}, \delta t\geq0,
\]
Hence, the basic reproduction number is given as,
\begin{equation}
R_0
=
\frac{\langle k_{in}k_{out}\rangle}
{\langle k_{in}\rangle}
\left[
1-
\left(
1+\frac{\delta t}{t_0}
\right)^{-(\alpha-2)}
\right].
\end{equation}
\begin{figure}
    \centering
    \includegraphics[width=\linewidth]{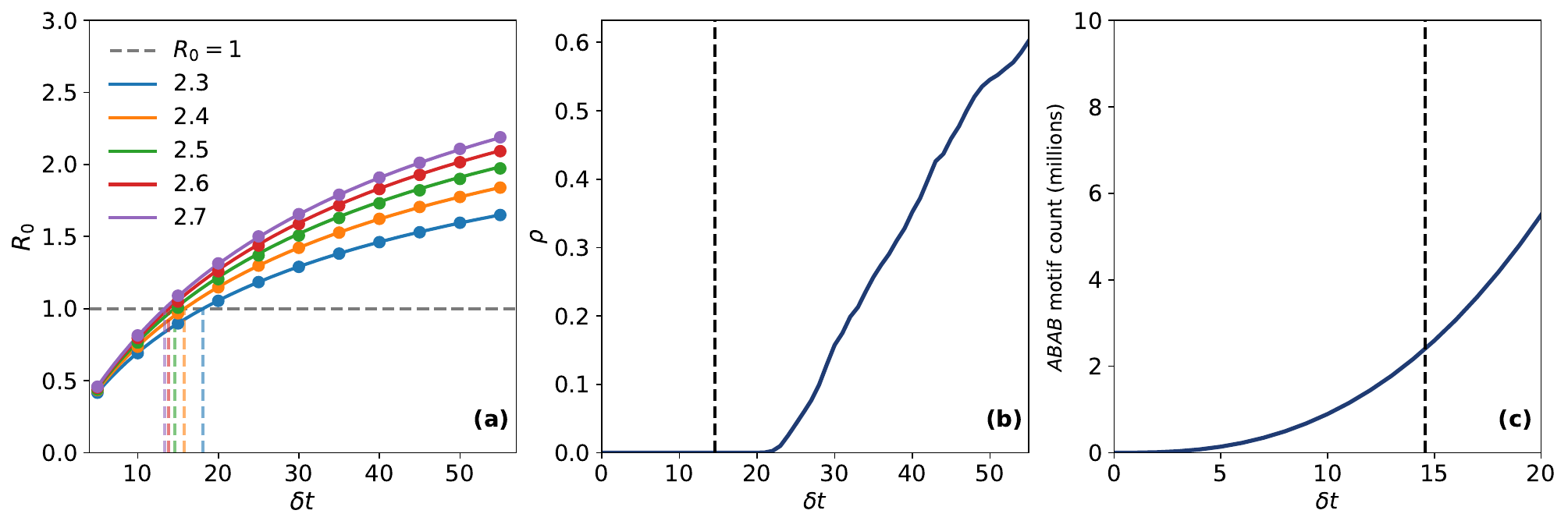}
    \caption{
    Undirected frSIS spreading on  \(k=3\) regular temporal networks with
    Lomax-distributed link inter-event times.
    (a) Analytical \(R_0\) curves with numerical measurements (symbols) for different $\alpha$ (colours).
    (b) Steady state density \(\rho\) for \(\alpha=2.5\). The dashed line marks the analytically predicted transition point.
    (c) Number of alternating
    \(A\rightarrow B\rightarrow A\rightarrow B\) temporal paths, shown in millions. 
    }
    \label{fig:undirected}
\end{figure}

\section{Extension to undirected temporal networks}
\label{app:undirected}
The proposed methods extend naturally to undirected temporal networks. For example, in the case of the frSIS process, similarly to the directed case, spreading may occur from an event to the first admissible event on each adjacent link, in this case excluding the focal link. As earlier, we can describe this as a branching process via generating function theory. Let $p_k$ be the degree distribution of the underlying, undirected static network, and $F_R(\delta t)$ be the cumulative residual function of the inter-event time distribution on each link. The generating function for the excess degree distribution of the resulting event graph is therefore
\begin{equation}
    G_1(y)=\left(\sum_{k=0}\frac{k p_k}{\langle k\rangle}H(y)^{k-1}\right)^2,
\end{equation}
where $H(y)=1-F_R(\delta t)+F_R(\delta t)y$.

The basic reproduction number is therefore given by the mean excess out-degree
\begin{equation}
    R_0=G_1'(1)=2F_R(\delta t)\frac{\langle k^2 \rangle-\langle k \rangle}{\langle k \rangle}.
\end{equation}
In k-regular networks, this reduces to 
\begin{equation}
    R_0=2(k-1)F_R(\delta t).
\end{equation}

As shown in Fig.~\ref{fig:undirected}(a), this expression accurately predicts the local branching factor. However, the condition significantly underestimates the onset of the frSIS process, as we can see in Fig.~\ref{fig:undirected}(b). This mismatch arises because the branches in the event graph generated from undirected temporal networks are not independent. In particular, paths in the event graph can alternate between just two physical links,
\begin{equation*}
    (a,b) \rightarrow (b,c) \rightarrow(a,b) \rightarrow (b,c),
\end{equation*}
which we call $ABAB$ motifs. These motifs cause branches to converge rather than opening up new distinct spreading pathways, hence shifting the predicted threshold below what is observed, as we can see in Fig~\ref{fig:undirected}(c). Despite the breakdown of precise analytical predictions, the analytical expression can still capture the effects of temporal and static structure at a coarse grain level, and the numerical advantages of the event graph approach are retained. In principle, the $ABAB$ motifs could be removed by preventing paths from returning the infection source, however, the event graph structure would then depend on the process history and would hence no longer be globally consistent.

\end{document}